\documentclass[conference]{IEEEtran}
\IEEEoverridecommandlockouts
\usepackage{graphicx} 
\graphicspath{{./figures}}
\usepackage[OT1]{fontenc}
\usepackage{color}
\usepackage[dvipsnames,svgnames, table]{xcolor} 
\usepackage{siunitx}
\usepackage{booktabs}
\usepackage[backend=biber,style=ieee]{biblatex}

\usepackage{tabularx}
\usepackage{tablefootnote}
\usepackage{pgfplotstable}
\usetikzlibrary{spy}
\usepgfplotslibrary{colorbrewer}

\AtBeginBibliography{\footnotesize}

\usepackage{xifthen}
\newcommand{\prob}[1][]{
\ifthenelse{\isempty{#1}}%
      {\ensuremath{P}}%
    {\ensuremath{P\left\(#1\right\)}}%
}

\usepackage{amsmath,amssymb,bm}

\usepackage[dvipsnames,svgnames, table]{xcolor} 

\usepackage{listofitems}
\usepackage{pgffor}

\makeatletter

\def\myColorList{LimeGreen,BrickRed,Fuchsia,Bittersweet,YellowOrange, YellowGreen, WildStrawberry}
\readlist\ColorList\myColorList

\newcommand{\defineauthors}[1]{

    \foreach \x [count=\xi from 1] in {#1} { 
    \expandafter\xdef\csname\x\endcsname####1{\noexpand\textcolor{\ColorList[\xi]}{[\unexpanded\expandafter{\x}: ####1]}}
    }
  
}
\makeatother

\usepackage{array}
\newcolumntype{H}{>{\setbox0=\hbox\bgroup}c<{\egroup}@{}}

\usepackage[font={small}]{caption}
\usepackage{subcaption}
\usepackage[capitalize]{cleveref}
\usepackage{xspace}

\usepackage{placeins}
\usepackage[xindy, nonumberlist]{glossaries}
\makenoidxglossaries

\newacronym{2d}{2D}{two-dimensional}
\newacronym{3d}{3D}{three-dimensional}
\newacronym{3gpp}{3GPP}{3rd generation partnership project}
\newacronym{4g}{4G}{fourth-generation}
\newacronym{5g}{5G}{fifth-generation}
\newacronym{6g}{6G}{sixth-generation}
\newacronym{abp}{ABP}{authentication by personalisation}
\newacronym{ac}{AC}{alternating current}
\newacronym{aclr}{ACLR}{adjacent channel leakage ratio}
\newacronym{adc}{ADC}{analog-to-digital converter}
\newacronym{adr}{ADR}{adaptive data rate}
\newacronym{afe}{AFE}{analog front end}
\newacronym{ag}{AG}{array gain}
\newacronym{ai}{AI}{artificial intelligence}
\newacronym{am}{AM}{amplitude modulation}
\newacronym{amam}{AM/AM}{amplitude modulation to amplitude modulation}
\newacronym{amp}{AMP}{approximate message passing}
\newacronym{ampm}{AM/PM}{amplitude modulation to phase modulation}
\newacronym{aoa}{AOA}{angle-of-arrival}
\newacronym{aod}{AOD}{angle-of-departure}
\newacronym{ap}{AP}{access point}
\newacronym{api}{API}{application program interface}
\newacronym{apu}{APU}{access point unit}
\newacronym{ar}{AR}{augmented reality}
\newacronym{arp}{ARP}{Antenna Reference Point}
\newacronym{asic}{ASIC}{application specific integrated circuit}
\newacronym{ask}{ASK}{amplitude-shift keying}
\newacronym{auv}{AUV}{autonomous underwater vehicle}
\newacronym{awgn}{AWGN}{additive white Gaussian noise}
\newacronym{baw}{BAW}{bulk acoustic wave}
\newacronym{bb}{BB}{base-band}
\newacronym{bc}{BC}{backscatter communication}
\newacronym{bd}{BD}{backscatter device}
\newacronym{ber}{BER}{bit error rate}
\newacronym{bf}{BF}{beamforming}
\newacronym{bldc}{BLDC}{brushless DC}
\newacronym{bom}{BOM}{bill of materials}
\newacronym{bs}{BS}{base station}
\newacronym{bw}{BW}{bandwidth}
\newacronym{ca}{CA}{carrier emitter}
\newacronym{cad}{CAD}{channel activity detection}
\newacronym{cars}{CARS}{calibration reference signal}
\newacronym{cbm}{CBM}{condition based maintenance}
\newacronym{cc}{CC}{constant current}
\newacronym{ccdf}{CCDF}{complementary cumulative distribution function}
\newacronym{ccnn}{CCNN}{circular convolutional neural network}
\newacronym{ccs}{CCS}{correlative channel sounder}
\newacronym{cdf}{CDF}{cumulative distribution function}
\newacronym{cdrx}{CDRX}{connected mode DRX}
\newacronym{ce}{CE}{coverage enhancement}
\newacronym{ced}{CED}{cumulative energy density}
\newacronym{cf}{CF}{cell-free}
\newacronym{cfmmimo}{CF-mMIMO}{cell-free massive MIMO}
\newacronym{cfo}{CFO}{carrier frequency offset}
\newacronym{cir}{CIR}{channel impulse response}
\newacronym{cmos}{CMOS}{complementary metal oxide semiconductor}
\newacronym{cost}{COST}{commercial off-the-shelf}
\newacronym{cots}{COTS}{commercial off-the-shelf}
\newacronym{cp}{CP}{cyclic prefix}
\newacronym{cpt}{CPT}{capacitive power transfer}
\newacronym{cpu}{CPU}{central-processing unit}
\newacronym{cqi}{CQI}{channel quality indicator}
\newacronym{cr}{CR}{coding rate}
\newacronym{crc}{CRC}{cyclic redundancy check}
\newacronym{crlb}{CRLB}{Cram\'er-Rao lower bound}
\newacronym{crs}{CRS}{cell reference signal}
\newacronym{cs}{CS}{compressed sensing}
\newacronym{cse}{CSE}{channel state estimation}
\newacronym{csi}{CSI}{channel state information}
\newacronym{csp}{CSP}{contact service point}
\newacronym{css}{CSS}{chirp spread spectrum}
\newacronym{cu}{CU}{central unit}
\newacronym{cv}{CV}{constant voltage}
\newacronym{cw}{CW}{continuous wave}
\newacronym{dab}{DAB}{Digital Audio Broadcasting}
\newacronym{dac}{DAC}{digital-to-analog converter}
\newacronym{daq}{DAQ}{data acquisition system}
\newacronym{das}{DAS}{distributed antenna systems}
\newacronym{dc}{DC}{direct current}
\newacronym{dcc}{DCC}{dynamic cooperation clustering}
\newacronym{de}{DE}{drain efficiency}
\newacronym{dft}{DFT}{discrete Fourier transform}
\newacronym{dl}{DL}{downlink}
\newacronym{dlc}{DLC}{Distributed Laser Charging}
\newacronym{dli}{DLI}{direct link interference}
\newacronym{dm}{DM}{diffuse multipath}
\newacronym{dma}{DMA}{Direct Memory Access}
\newacronym{dmc}{DMC}{diffuse multipath component}
\newacronym{dmimo}{D-MIMO}{distributed MIMO}
\newacronym{doa}{DOA}{direction-of-arrival}
\newacronym{dpd}{DPD}{digital pre-distortion}
\newacronym{dpdk}{DPDK}{Data Plane Development Kit}
\newacronym{dram}{DRAM}{dynamic random-access memory}
\newacronym{drx}{DRX}{Discontinuous Reception Mode}
\newacronym{dsl}{DSL}{digital subscriber line}
\newacronym{dsp}{DSP}{digital signal processing}
\newacronym{dss}{DSS}{Dataset Storage Standard}
\newacronym{duc}{DUC}{digital up-converter}
\newacronym{dvb}{DVB}{Digital Video Broadcasting}
\newacronym{e2e}{E2E}{end-to-end}
\newacronym{easa}{EASA}{European Union Aviation Safety Agency}
\newacronym{ec}{EC}{European Commission}
\newacronym{ecdf}{eCDF}{empirical cumulative distribution function}
\newacronym{ecsp}{ECSP}{edge computing service point}
\newacronym{edlc}{EDLC}{electrostatic double-layer capacitors}
\newacronym{edrx}{eDRX}{Extended Discontinuous Reception Mode}
\newacronym{ee}{EE}{energy efficiency}
\newacronym{egprs}{EGPRS}{Enhanced Data Rates for GSM Evolution}
\newacronym{eh}{EH}{energy harvesting}
\newacronym{eirp}{EIRP}{equivalent isotropically radiated power}
\newacronym{em}{EM}{electromagnetic}
\newacronym{embb}{eMBB}{enhanced Mobile Broadband}
\newacronym{en}{EN}{energy neutral}
\newacronym{end}{END}{energy neutral device}
\newacronym{enob}{ENOB}{effective number of bits}
\newacronym{eol}{EoL}{end of life}
\newacronym{epu}{EPU}{edge processing unit}
\newacronym{erp}{ERP}{equivalent radiated power}
\newacronym[plural=ESCs,firstplural=electronic speed controllers (ESCs)]{esc}{ESC}{electronic speed control}
\newacronym{esd}{ESD}{electrostatic discharge}
\newacronym{esl}{ESL}{electronic shelf label}
\newacronym{esr}{ESR}{equivalent series resistance}
\newacronym{etsi}{ETSI}{European Telecommunications Standards Institute}
\newacronym{evd}{EVD}{eigenvalue decomposition}
\newacronym{evm}{EVM}{Error Vector Magnitude}
\newacronym{fa}{FA}{federation anchor}
\newacronym{fair}{FAIR}{Findability, Accessibility, Interoperability, and Reuse of digital assets}
\newacronym{fd}{FD}{front-haul distance}
\newacronym{fdma}{FDMA}{frequency division multiple access}
\newacronym{fem}{FEM}{finite element analysis}
\newacronym{fembb}{feMBB}{further enhanced mobile broadband}
\newacronym{fft}{FFT}{fast Fourier transform}
\newacronym{fh}{FH}{fronthaul}
\newacronym{fim}{FIM}{Fisher information matrix}
\newacronym{fits}{FITS}{Flexible Image Transport System}
\newacronym{fom}{FoM}{figure of merit}
\newacronym{fpga}{FPGA}{field-programmable gate array}
\newacronym{fsk}{FSK}{frequency shift keying}
\newacronym{fss}{FSS}{frequency selective surface}
\newacronym{gb}{GB}{grant-based}
\newacronym{gf}{GF}{grant-free}
\newacronym{gnb}{gNB}{Next Generation Node B}
\newacronym{gnn}{GNN}{graph neural network}
\newacronym{gnss}{GNSS}{global navigation satellite system}
\newacronym{gpclk}{GPCLK}{general purpose clock}
\newacronym{gprs}{GPRS}{General Packet Radio Services}
\newacronym{gps}{GPS}{Global Positioning System}
\newacronym{gpu}{GPU}{graphical processing unit}
\newacronym{gscm}{GSCM}{geometry‐based stochastic model}
\newacronym{gsm}{GSM}{Global System for Mobile Communications}  %
\newacronym{gwp}{GWP}{Global Warming Potential}
\newacronym{harq}{HARQ}{hybrid automatic repeat request}
\newacronym{hat}{HAT}{hardware attached on top}
\newacronym{hcs}{HCS}{human-centric services}
\newacronym{hdf5}{HDF5}{Hierarchical Data Format version 5}
\newacronym{hpbm}{HPBM}{half power beam width}
\newacronym{HyMPRo}{HyMPRo}{Hybrid Multi-Path Routing algorithm}
\newacronym{i.i.d.}{i.i.d.}{independent and identically distributed}
\newacronym{ib}{IB}{in-band}
\newacronym{ibo}{IBO}{input back-off}
\newacronym{ic}{IC}{integrated circuit}
\newacronym{id}{ID}{information decoding}
\newacronym{idft}{IDFT}{inverse discrete Fourier transform}
\newacronym{if}{IF}{intermediate-frequency}
\newacronym{iid}{i.i.d.}{independently and identically distributed}
\newacronym{iis}{IIS}{integrated information system}
\newacronym{im}{IM}{intermodulation}
\newacronym{imu}{IMU}{inertial measurement unit}
\newacronym{io}{IO}{input/output}
\newacronym{iot}{IoT}{Internet of Things}
\newacronym{ipt}{IPT}{inductive power transfer}
\newacronym{ipy}{IPY}{Interventions per Year}
\newacronym{iq}{IQ}{in-phase and quadrature}
\newacronym{ir}{IR}{infrared}
\newacronym{ici}{ICI}{intercarrier interference}
\newacronym{isi}{ISI}{intersymbol interference}
\newacronym{ism}{ISM}{industrial, scientific and medical}
\newacronym{isp}{ISP}{internet service provider}
\newacronym{jfet}{JFET}{junction field effect transistor}
\newacronym{kpi}{KPI}{key performance indicator}
\newacronym{kvi}{KVI}{key value indicator}
\newacronym{larva}{LARVA}{LARge Virtual Array}
\newacronym{lca}{LCA}{life cycle assessment}
\newacronym{lco}{LCO}{lithium cobalt oxide}
\newacronym{ldo}{LDO}{Low-dropout voltage regulator}
\newacronym{ldpc}{LDPC}{low-density parity-check}
\newacronym{less}{LESS}{Low Energy Scheduler Solution}
\newacronym{lfp}{LFP}{lithium iron phosphate}
\newacronym{lic}{LIC}{lithium-ion capacitor}
\newacronym{lidar}{LiDAR}{light detection and ranging}
\newacronym{liion}{Li-ion}{lithium-ion}
\newacronym{lipo}{LiPo}{lithium polymer}
\newacronym{lis}{LIS}{large intelligent surface}
\newacronym{llh}{LLH}{log-likelihood}
\newacronym{lmmse}{LMMSE}{least minimum mean square error}
\newacronym{lmo}{LMO}{lithium ion manganese oxide}
\newacronym{lna}{LNA}{low-noise amplifier}
\newacronym{lo}{LO}{local oscillator}
\newacronym{lora}{LoRa}{long range}
\newacronym{lorawan}{LoRaWAN}{long-range wide-area network}
\newacronym{los}{LoS}{line-of-sight}
\newacronym{lp}{LP}{linear programming}
\newacronym{lpt}{LPT}{laser power transfer}
\newacronym{lpwa}{LPWA}{Low Power Wide Area}
\newacronym{lpwan}{LPWAN}{low-power wide-area network}
\newacronym{lqi}{LQI}{link quality indicator}
\newacronym{lrelu}{LReLU}{leaky rectified linear unit}
\newacronym{lrt}{LRT}{likelihood-ratio test}
\newacronym{ls}{LS}{least squares}
\newacronym{lsa}{LSA}{large synthetic array}
\newacronym{lsf}{LSF}{large-scale fading}
\newacronym{lsfc}{LSFC}{large-scale fading component}
\newacronym{ltc}{LTC}{lithium thionyl chloride}
\newacronym{lte}{LTE}{Long Term Evolution}
\newacronym{lti}{LTI}{linear time-invariant}
\newacronym{lto}{LTO}{lithium titanate}
\newacronym{m2m}{M2M}{machine to machine}
\newacronym{mac}{MAC}{Medium Access Control}
\newacronym{mate}{MATE}{millimeter-wave MIMO testbed}
\newacronym{mc}{MC}{Monte Carlo}
\newacronym{mcl}{MCL}{Maximum Coupling Loss}
\newacronym{mcs}{MCS}{modulation and coding scheme}
\newacronym{mcu}{MCU}{Microcontroller Unit}
\newacronym{mec}{MEC}{multi-access edge computing}
\newacronym{mems}{MEMS}{micro-electromechanical systems}
\newacronym{mf}{MF}{matched filter}
\newacronym{mimo}{MIMO}{multiple-input multiple-output}
\newacronym{miso}{MISO}{multiple-input single-output}
\newacronym{ml}{ML}{machine learning}
\newacronym{mlp}{MLP}{multilayer perceptron}
\newacronym{mmimo}{mMIMO}{massive MIMO}
\newacronym{mmse}{MMSE}{minimum mean square error}
\newacronym{mmtc}{mMTC}{massive machine-typed communication}
\newacronym{mmwave}{mmWave}{millimeter wave}
\newacronym{mn}{MN}{matching network}
\newacronym{mosfet}{MOSFET}{metal-oxide semiconductor field effect transistor}
\newacronym{mpc}{MPC}{multipath component}
\newacronym{mppt}{MPPT}{maximum power point tracking}
\newacronym{mr}{MR}{maximum ratio}
\newacronym{mrc}{MRC}{maximum ratio combining}
\newacronym{mrc_em}{MRC}{maximum ratio combining}
\newacronym{mrc_EM}{MRC}{Magnetic Resonance Coupling}
\newacronym{mrt}{MRT}{maximum ratio transmission}
\newacronym{mse}{MSE}{mean square error}
\newacronym{mtc}{MTC}{Machine-Type Communication}
\newacronym{multi-rat}{Multi-RAT}{multiple radio access technology}
\newacronym{music}{MUSIC}{MUltiple SIgnal Classification}
\newacronym{nb}{NB}{narrowband}
\newacronym{nbiot}{NB-IoT}{narrowband IoT}
\newacronym{nca}{NCA}{nickel cobalt aluminum}
\newacronym{netcdf}{NetCDF}{Network Common Data Form}
\newacronym{nfc}{NFC}{near-field communication}
\newacronym{ni}{NI}{National Instruments}
\newacronym{nicd}{NiCd}{nikkel cadmium}
\newacronym{nimh}{NiMH}{nikkel metal hydride}
\newacronym{nlos}{NLoS}{non-line-of-sight}
\newacronym{nmc}{NMC}{nickel manganese cobalt}
\newacronym{nmos}{nMOS}{n-channel metal-oxide semiconductor}
\newacronym{nn}{NN}{neural network}
\newacronym{nnls}{NNLS}{non-negative least squares}
\newacronym{noma}{NOMA}{non-orthogonal multiple access}
\newacronym{np}{NP}{Neyman-Pearson}
\newacronym{npbch}{NPBCH}{Narrowband Physical Broadcast Channel}
\newacronym{npss}{NPSS}{Narrow Band Primay Synchronization Signal}
\newacronym{nr}{NR}{New Radio}
\newacronym{nrs}{NRS}{Narrow Band Reference Signal}
\newacronym{nsss}{NSSS}{Narrowband Secondary Synchronization Signal}
\newacronym{ntp}{NTP}{network time protocol}
\newacronym{oai}{OAI}{OpenAirInterface} 
\newacronym{ofdm}{OFDM}{orthogonal frequency-division multiplexing}
\newacronym{ofdmim}{OFDM-IM}{OFDM with index modulation}
\newacronym{oob}{OOB}{out-of-band}
\newacronym{ook}{OOK}{on-off keying}
\newacronym{oran}{O-RAN}{open radio-access network}
\newacronym{os}{OS}{operating system}
\newacronym{ota}{OTA}{over-the-air}
\newacronym{otaa}{OTAA}{over-the-air authentication}
\newacronym{p1}{P1}{Phase 1}
\newacronym{p2}{P2}{Phase 2}
\newacronym{p2p}{P2P}{point-to-point}
\newacronym{pa}{PA}{power amplifier}
\newacronym{pae}{PAE}{power-added efficiency}
\newacronym{pana}{PanA}{Panel A}
\newacronym{panb}{PanB}{Panel B}
\newacronym{papr}{PAPR}{peak-to-average power ratio}
\newacronym{pc}{PC}{pilot count}
\newacronym{pcb}{PCB}{printed circuit board}
\newacronym{pcg}{PCG}{power consumption gain}
\newacronym{pcsi}{PCSI}{perfect channel state information}
\newacronym{pd}{PD}{powered device}
\newacronym{pdcch}{PDCCH}{physical downlink control channel}
\newacronym{pdf}{PDF}{probability density function}
\newacronym{pdp}{PDP}{power delay profile}
\newacronym{pdsch}{PDSCH}{physical downlink shared channel}
\newacronym{peb}{PEB}{positioning error bound}
\newacronym{per}{PER}{packet error rate}
\newacronym{pg}{PG}{path gain}
\newacronym{pgd}{PGD}{proximal gradient descent}
\newacronym{phy}{PHY}{physical}
\newacronym{pl}{PL}{path loss}
\newacronym{pll}{PLL}{phase-locked loop}
\newacronym{poe}{PoE}{power-over-Ethernet}
\newacronym{pps}{1PPS}{pulse per second}
\newacronym{prbs}{PRBs}{Physical Resource Blocks}
\newacronym{ps}{PS}{Processing System}
\newacronym{psd}{PSD}{power spectral density}
\newacronym{pse}{PSE}{power sourcing equipment}
\newacronym{psk}{PSK}{phase shift keying}
\newacronym{psm}{PSM}{power saving mode}
\newacronym{pss}{PSS}{primary synchronisation signal}
\newacronym{ptp}{PTP}{precision-time protocol}
\newacronym{ptrs}{PTRS}{Phase-Tracking Reference Signals}
\newacronym{ptw}{PTW}{paging time window}
\newacronym{pw}{PW}{planar wavefront}
\newacronym{pwm}{PWM}{pulse width modulation}
\newacronym{qam}{QAM}{quadrature amplitude modulation}
\newacronym{qos}{QoS}{quality-of-service}
\newacronym{quadriga}{QuaDRiGa}{QUAsi Deterministic RadIo channel GenerAtor}
\newacronym{ra}{RA}{Random Access}
\newacronym{ram}{RAM}{random-access memory}
\newacronym{ran}{RAN}{radio access network}
\newacronym{rar}{RAR}{Random Access Response}
\newacronym{rat}{RAT}{radio access technology}
\newacronym{rb}{Rb}{Rubidium}
\newacronym{rbs}{RBS}{radio base station}
\newacronym{rcs}{RCS}{radar cross section}
\newacronym{re}{RE}{radio element}
\newacronym{relu}{ReLU}{rectified linear unit}
\newacronym{rf}{RF}{radio frequency}
\newacronym{rfeh}{RFEH}{radio frequency energy harvesting}
\newacronym{rfic}{RFIC}{radio-frequency integrated circuit}
\newacronym{rfid}{RFID}{radio frequency identification}
\newacronym{rfpt}{RFPT}{radio frequency power transfer}
\newacronym{rfsoc}{RFSoC}{Radio Frequency System-on-Chip}
\newacronym{rir}{RIR}{room impulse response}
\newacronym{ris}{RIS}{reflective intelligent surface}
\newacronym{rllmtc}{RLLMTC}{reliable low latency machine type communication}
\newacronym{rls}{RLS}{recursive least squares}
\newacronym{rms}{RMS}{root mean square}
\newacronym{rmse}{RMSE}{root-mean-square error}
\newacronym{rmt}{RMT}{random matrix theory}
\newacronym{rof}{RoF}{radio-over-fiber}
\newacronym{ros}{ROS}{robot operating system}
\newacronym{rpi}{RPi}{Raspberry Pi}
\newacronym{rrc}{RRC}{Radio Resource Connection}
\newacronym{rreq}{RREQ}{route request packet}
\newacronym{rsrp}{RSRP}{Reference Signals Received Power}
\newacronym{rss}{RSS}{received signal strength}
\newacronym{rssi}{RSSI}{received signal strength indicator}
\newacronym{rtc}{RTC}{real time clock}
\newacronym{rtk}{RTK}{real time kinematics}
\newacronym{rts}{RTS}{ray tracing simulator}
\newacronym{rv}{RV}{random variable}
\newacronym{rw}{RW}{RadioWeaves}
\newacronym{rx}{RX}{receiver}
\newacronym{rzf}{RZF}{regularized zero forcing}
\newacronym{s-parameter}{S-parameter}{scattering parameter}
\newacronym{sa}{SA}{synchronization anchor}
\newacronym{sar}{SAR}{specific absorption rate}
\newacronym{scfdma}{SCFDMA}{single-carrier frequency division multiple access}
\newacronym{sdg}{SDG}{Sustainable Development Goal}
\newacronym{sdm}{SDM}{sigma-delta modulator}
\newacronym{sdn}{SDN}{software-defined network}
\newacronym{sdof}{SDoF}{sigma-delta over fiber}
\newacronym{sdr}{SDR}{software-defined radio}
\newacronym{sf}{SF}{spreading factor}
\newacronym{sfn}{SFN}{single frequency network}
\newacronym{sfp}{SFP}{small form-factor pluggable}
\newacronym{SigMF}{SigMF}{Signal Metadata Format}
\newacronym{simo}{SIMO}{single-input multiple-output}
\newacronym{sinr}{SINR}{signal-to-interference-plus-noise ratio}
\newacronym{siso}{SISO}{single-input single-output}
\newacronym{slam}{SLAM}{simultaneous localization and mapping}
\newacronym{slc}{SLC}{spatial leakage suppression}
\newacronym{sma}{SMA}{SubMiniature version A}
\newacronym{smc}{SMC}{specular multipath component}
\newacronym{smps}{SMPS}{switched mode power supply}
\newacronym{sndr}{SNDR}{signal-to-noise-and-distortion ratio}
\newacronym{snidr}{SNIDR}{signal-to-noise-and-interference-and-distortion ratio}
\newacronym{snr}{SNR}{signal-to-noise ratio}
\newacronym{soc}{SoC}{state of charge}
\newacronym{sram}{SRAM}{static random-access memory}
\newacronym{srd}{SRD}{short-range device}
\newacronym{srs}{SRS}{Sounding Reference Signal}
\newacronym{ssb}{SSB}{synchronisation signal block}
\newacronym{ssd}{SSD}{solid state drive}
\newacronym{steam}{STEAM}{science, technology, engineering, the arts, and mathematics}
\newacronym{svd}{SVD}{singular value decomposition}
\newacronym{sw}{SW}{spherical wavefront}
\newacronym{synce}{SyncE}{Synchronous Ethernet}
\newacronym{tau}{TAU}{tracking area update}
\newacronym{tcer}{TCER}{transported to consumed energy ratio}
\newacronym{tcp}{TCP}{Transmission Control Protocol}
\newacronym{tdd}{TDD}{time division duplexing}
\newacronym{tdoa}{TDOA}{time-difference-of-arrival}
\newacronym{to}{TO}{timing offset}
\newacronym{toa}{TOA}{time-of-arrival}
\newacronym{tof}{ToF}{time-of-flight}
\newacronym{tosm}{TOSM}{through-open-short-match}
\newacronym{trl}{TRL}{technology readyness level}
\newacronym{trp}{TRP}{Transmission Reception Point}
\newacronym{tsn}{TSN}{time-sensitive networking}
\newacronym{ttm}{TTM}{time to market}
\newacronym{ttn}{TTN}{The Things Network}
\newacronym{tx}{TX}{transmitter}
\newacronym{uav}{UAV}{unmanned aerial vehicle}
\newacronym{udp}{UDP}{User Datagram Protocol}
\newacronym{ue}{UE}{user equipment}
\newacronym{ugv}{UGV}{unmanned ground vehicle}
\newacronym{uhd}{UHD}{USRP hardware driver}
\newacronym{uhf}{UHF}{ultra-high frequency}
\newacronym{ul}{UL}{uplink}
\newacronym{ula}{ULA}{uniform linear array}
\newacronym{ummtc}{umMTC}{ultra massive machine type communication}
\newacronym{un}{UN}{United Nations}
\newacronym{upa}{UPA}{uniform planar array}
\newacronym{ura}{URA}{uniform rectangular array}
\newacronym{urllc}{URLLC}{ultra-reliable low-latency communications}
\newacronym{usrp}{USRP}{universal software radio peripheral}
\newacronym{uv}{UV}{unmanned vehicle}
\newacronym{uwb}{UWB}{ultrawideband}
\newacronym{v2v}{V2V}{vehicle-to-vehicle}
\newacronym{vep}{VEP}{virtual edge platform}
\newacronym{vlc}{VLC}{visible light communication}
\newacronym{vlp}{VLP}{visible light positioning}
\newacronym{vna}{VNA}{vector network analyzer}
\newacronym{votable}{VOTable}{Virtual Observatory Table}
\newacronym{vr}{VR}{virtual reality}
\newacronym{wb}{WB}{wideband}
\newacronym{wban}{WBAN}{wireless body area network}
\newacronym{wimax}{WiMAX}{Worldwide Interoperability for Microwave Access}
\newacronym{wlan}{WLAN}{wireless LAN}
\newacronym{wpt}{WPT}{wireless power transfer}
\newacronym{wr}{WR}{White Rabbit}
\newacronym{wrsn}{WRSN}{wireless rechargeable sensor network}
\newacronym{wsn}{WSN}{wireless sensor network}
\newacronym{xets}{XETS}{cross exponentially tapered slot}
\newacronym{xr}{XR}{extended reality}
\newacronym{z3ro}{Z3RO}{zero third-order distortion}
\newacronym{zf}{ZF}{zero-forcing}
\newacronym{zmcscg}{ZMCSCG}{zero mean circularly symmetric complex Gaussian}

\newacronym{bpsk}{BPSK}{binary phase-shift keying}
\newacronym{qpsk}{QPSK}{quadrature phase-shift keying}

\newacronym{ser}{SER}{symbol-error rate}
\newacronym{pcie}{PCIe}{Peripheral Component Interconnect Express}

\usepackage{todonotes}

\usepackage{comment}

\newcolumntype{R}{>{\raggedleft\arraybackslash}X}

\title{Energy Reduction in Cell-Free Massive MIMO through Fine-Grained Resource Management
\thanks{The work by \"O. T. Demir was supported by 2232-B International Fellowship for Early Stage Researchers Programme funded by the Scientific and Technological Research Council of T\"urkiye. The work by L. Méndez is supported by the Spanish National project IRENE-EARTH (PID2020-115323RB-C33/AEI/10.13039/501100011033) and PASSIONATE under the CHIST-ERA grant CHIST-ERA-22-WAI-04, by AEI PCI2023-145990-2. The work by G.~Callebaut is supported by the REINDEER project under grant agreement No.~101013425. This work has partly been funded by the Excellence Center Linköping – Lund in Information Technology (ELLIIT).}
}

\author{
    \IEEEauthorblockN{%
    Özlem Tuğfe Demir\IEEEauthorrefmark{1}, 
    Lianet Méndez-Monsanto\IEEEauthorrefmark{4}, 
    Nicola Bastianello\IEEEauthorrefmark{2}, 
    Emma Fitzgerald\IEEEauthorrefmark{3}, 
    Gilles Callebaut\IEEEauthorrefmark{5}}%
    \IEEEauthorblockA{\IEEEauthorrefmark{1} \textit{TOBB University of Economics and Technology}, Ankara, Türkiye}

\IEEEauthorblockA{\IEEEauthorrefmark{4} \textit{Department of Signal Theory and Communications, Universidad Carlos III de Madrid}, Madrid, Spain}

    \IEEEauthorblockA{\IEEEauthorrefmark{2} \textit{KTH Royal Institute of Technology}, Sweden}
    \IEEEauthorblockA{\IEEEauthorrefmark{3} \textit{Department of Electrical and Information Technology, Lund University}, Lund, Sweden}

    \IEEEauthorblockA{\IEEEauthorrefmark{5} \textit{Department of Electrical Engineering, KU Leuven}, Belgium}

}

\begin{document}

\maketitle

\begin{abstract}
The physical layer foundations of \gls{cfmmimo} have been well-established. As a next step, researchers are investigating practical and energy-efficient network implementations. This paper focuses on multiple sets of \glspl{ap} where \glspl{ue} are served in each set, termed a federation, without inter-federation interference. The combination of federations and \gls{cfmmimo} shows promise for highly-loaded scenarios. Our aim is to minimize the total energy consumption while adhering to \gls{ue} downlink data rate constraints. The energy expenditure of the full system is modelled using a detailed hardware model of the \glspl{ap}. We jointly design the AP-UE association variables, determine active \glspl{ap}, and assign \glspl{ap} and \glspl{ue} to federations. To solve this highly combinatorial problem, we develop a novel alternating optimization algorithm. Simulation results for an indoor factory demonstrate the advantages of considering multiple federations, particularly when facing large data rate requirements. Furthermore, we show that adopting a more distributed \gls{cfmmimo} architecture is necessary to meet the data rate requirements. Conversely, if feasible, using a less distributed system with more antennas at each \gls{ap} is more advantageous from an energy savings perspective.
\end{abstract}

\glsresetall 
\section{Introduction}\label{sec:introduction}
One particularly promising advancement in 6G is the adoption of \gls{cfmmimo}. In this architectural paradigm, \glspl{ap} equipped with single or multiple antennas are strategically dispersed across a defined geographical area \cite{8097026}. Unlike the conventional cellular paradigm, where a \gls{ue} is exclusively served by a single AP, the essence of \gls{cfmmimo} lies in the concept of having multiple APs available to the UEs. Recent research endeavors have delved into user-centric \gls{cfmmimo}, wherein a network selects a subset of \glspl{ap} to serve a UE \cite{8761828,demir2021foundations}.

\begin{figure}[tb!]
    \centering
    \includegraphics[width=0.8\linewidth]{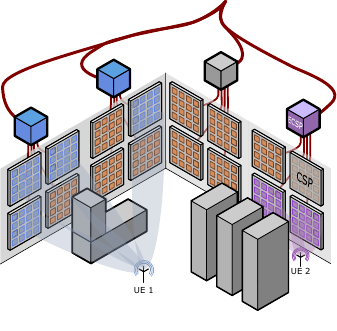}
    \caption{Illustration of the system with two \glspl{ue}, four \glspl{ecsp} and four \glspl{csp} coordinated by one \gls{ecsp}. Only a subset of resources (a federation) is activated during joint downlink precoding based on the \gls{ue} requirements and the propagation channel conditions. The  objective is to optimize the total energy expenditure of the network. Two federations are depicted, serving \gls{ue}~1 (blue) and \gls{ue}~2 (purple).}\label{fig:system}
    \vspace{-2mm}
\end{figure}

This paper contributes to the ongoing discourse by extending the paradigm of user-centric \gls{cfmmimo} to incorporate the energy consumption of the radio access network. While previous studies have primarily focused on maximizing the overall network capacity, our work takes a novel energy reduction approach by concentrating on the selection of a subset of \glspl{ap} in disjoint groups, where the objective is to meet \gls{ue} requirements, specifically focusing on the requested download data rate. This approach is illustrated in~\cref{fig:system}. A selection of radio resources where the resources are all serving the same set of \glspl{ue} is coined a federation. The APs are primarily represented by \glspl{csp}. A subset of CSPs is coordinated by a single \gls{ecsp}, which is the central processing unit (CPU) equivalent in CF-mMIMO terminology. The UEs within different federations are allocated orthogonal resources so different federations do not create interference to each other. This approach, together with active CSP and ECSP selection, provides an energy-efficient method to meet the high data rate requirements in an environment with densely deployed UEs.

\subsection{State-of-the-Art}

The research on \gls{ee} in \gls{cfmmimo} has yielded valuable insights, with a focus on three distinct objectives: 1) minimizing the energy consumption or maximizing the \gls{ee}, 2) enhancing performance, and 3) tackling scalability challenges associated with \gls{cfmmimo}.

In~\cite{8097026}, the authors present a closed-form result enabling the analysis of backhaul power consumption, the number of \glspl{ap}, and the number of antennas per \gls{ap} on total EE. Additionally, they design an optimal power allocation algorithm aiming to maximize total EE, subject to per-UE spectral efficiency and per-AP power constraints. Compared to equal power control, the proposed power allocation scheme doubles total EE. Two \gls{ap} selection schemes are proposed based on the received power and the largest large-scale fading to reduce power consumption caused by fronthaul links. Similar to our study, the study also focuses on downlink payload data transmission. They employ a power consumption model detailed in~\cite{7031971}. The scalability \textit{issues} of \gls{cfmmimo} are addressed in~\cite{8761828}. The authors of \cite{8761828} propose and evaluate an \gls{ap} selection method related to data processing, network topology, and power control, achieving full scalability with a modest performance loss compared to the canonical form of \gls{cfmmimo}. \glspl{ap} are grouped in cell-centric clusters, each managed by a CPU, leading to reduced deployment complexity. This is similar to our approach as detailed in~\cref{sec:model}. Other works investigating \gls{ap} selection strategies are~\cite{9860684, 9343832,9685221}, all inspired by the same approach, i.e., selection of near \glspl{ap}, deactivating far \glspl{ap} to avoid wasting power over channels with severe propagation losses.

In~\cite{demir2024}, the authors address the end-to-end power consumption in a \gls{cfmmimo} system on O-RAN architecture. The joint orchestration of cloud processing, fronthaul resources, and radio power is achieved through quality-of-service-aware and sum rate maximization optimization problems. In~\cite{chien2020}, the authors propose an energy-efficient load balancing strategy by minimizing total downlink power consumption at the \glspl{ap}, considering both transmit powers and hardware dissipation. The globally optimal solution is obtained by solving a mixed-integer second-order cone program. The proposed optimization algorithms significantly reduce power consumption compared to keeping all APs turned on, as validated using the energy model from~\cite{7031971}.

\subsection{Contributions}

This work presents a comprehensive hardware and energy model for \gls{cfmmimo} systems. We define a problem and propose a tailored divide-and-conquer heuristic aimed at minimizing total energy consumption while adhering to \gls{ue} data rate constraints and federation assignment rules. Importantly, our solution is open-source, promoting transparency and collaboration through the provided link to the source code on GitHub\footnote{\url{https://github.com/ozlemtugfedemir/cell-free-federations}}. We conduct a thorough evaluation on realistic use cases, specifically an indoor factory scenario. Our findings highlight the need for a paradigm shift, advocating for the optimization of energy consumption rather than traditional emphasis on spectral efficiency. This shift yields markedly different results, underscoring the significance of considering energy expenditure as a primary metric in system optimization. We demonstrate that more energy can be saved by using orthogonal resource sharing via federations when data rate requirements are demanding.

\section{Terminology, Concepts, and Models}\label{sec:model}\FloatBarrier

The network architecture considered in this work, adopted from~\cite{Call2207:Dynamic}, is depicted in~\cref{fig:system}. The currently employed techniques in cellular networks to distribute synchronization, data, and control cannot be scaled to accommodate for the envisioned \gls{cfmmimo} systems~\cite{callebautTestbeds}. Therefore, the system under study utilizes Ethernet to distribute synchronization, data, and control to a large amount of \glspl{ap}.

A distinction is made between \glspl{ap} providing the wireless access to the network and the \glspl{ap} coordinating several of these wireless \glspl{ap}. To differentiate between those two, we follow the terminology and concepts from~\cite{Call2207:Dynamic}, where the wireless \gls{ap} is called the \gls{csp} and the coordinator, the \gls{ecsp}. In this work, we have focused on \gls{dl} communication and thus the considered hardware blocks of the \glspl{csp} are: the \gls{pa}, fronthaul, sampling, and processing blocks; detailed below.

\subsection{Hardware and Energy Models}

A simplified hardware model is used in this work, to be able to optimize the energy consumption of the \gls{cfmmimo} system. An overview of the hardware blocks and their energy expenditure is summarized in~\cref{tab:energy}. The \glspl{csp} are coordinated by one or more \glspl{ecsp}. In this work, the model assumes that the \glspl{ecsp} are only responsible for the data transfer and thus only the energy consumption of data aggregation is included.

\begin{table*}[hbtp]
    \centering
    \caption{Considered energy consumption model and parameters of the system.}
    \label{tab:energy}
    \begin{tabularx}{\linewidth}[t]{@{}R l l l l R l l l l@{}}
    \cmidrule[\heavyrulewidth](r){1-5} \cmidrule[\heavyrulewidth](l){6-10}
     & Symbol & Expression &  Unit & Ref. & & Symbol & Value &  Unit & Ref.\\
    \cmidrule(r){1-5} \cmidrule(l){6-10}
    \multicolumn{1}{l}{\textbf{Component}} & & & & & \multicolumn{1}{l}{\textbf{Parameter}}\\
        Power amplifier & $P_\text{PA}$&  $1/\eta_\text{max} \sqrt{P_\text{t,max}}\sqrt{P_\text{t}}$ &\si{\watt}&  \cite{6515206} & Coherence time &$\tau_c$& 200&\si{symbols}\\
        \Acrshort{dac}/ADC & $P_\text{DAC}$ & $\text{FoM}_{W}\  2^{b} f_s$ &\si{\watt} &\cite{7569654} & Pilot time &$\tau_p$& &\si{symbols}\\
        \Acrlong{wr} core & $P_\text{sync}$ & 2.2 &\si{\watt}& \cite{Wrrefere77:online} &\Acrshort{rf} sampling rate & $f_s$ & \num{600} & \si{\mega\hertz} &\cite{8551268}\\ 
        Ethernet link & $P_\text{eth}$ & 7$^\ddag$  &\si{\watt}& \cite{9815696} & \Acrlong{bb} sampling rate & $f_\text{BB}$ & \num{20} & \si{\mega\hertz}\\
        \multicolumn{1}{l}{\textbf{Process}}& &&&& Max. \gls{pa} efficiency & $\eta_\text{max}$ &$0.34^*$&& \cite{powerAmplifierQorvo}\\
        \Acrshort{dsp} operation & $E_\text{op}$ & $\zeta (E_\text{mac}+\alpha E_\text{sram}$ &&&  Max. transmit power & $P_\text{t,max}$& \num{3} & \si{\watt} &\cite{powerAmplifierQorvo}\\
       && \quad$+\gamma E_\text{dram})$ &\si{\joule\per oper.} &  & &$E_\text{mac}$ & \num{3.1} & \si{\pico\joule\per oper.}& \cite{han2016eie}
        \\
        Channel estimation & $E_\text{CE}$ & $2MK E_\text{op} \tau_p$ &\si{\joule}& \cite{7569654} & &$E_\text{sram}$& \num{5}& \si{\pico\joule\per oper.}& \cite{han2016eie}\\
        Linear processing & $E_\text{LP}$ & $2MK E_\text{op} (\tau_c - \tau_p)$ &\si{\joule}& \cite{7569654}& &$E_\text{dram}$& \num{640} & \si{\pico\joule\per oper.}& \cite{han2016eie}\\
        DL transmit &$E_\text{PA}$& $ P_\text{PA} (\tau_c-\tau_p) f_\text{BB}^{-1}$ &\si{\joule} & & Energy overhead &$\zeta$ & 1.2\\
        \Acrshort{dac}/ADC energy &$E_\text{DAC}$& $M P_\text{DAC} (\tau_c f_\text{BB}^{-1})$ &\si{\joule} & &  \Acrshort{fom} (\acrshort{dac})&  $\text{FoM}_{W}$ & \num{34.4}& \si{fJ/step} & \cite{8551268}  \\
        \multicolumn{1}{l}{\textbf{Node}} & & & & & &  $\alpha$ & \SI{10}{\percent} && \cite{han2016eie}\\
        ECSP&$E_{\text{ECSP}}$& $(P_\text{eth}+P_\text{sync})  \tau_c f_\text{BB}^{-1}$& \si{\joule} & & & $\gamma$ & \SI{1}{\percent} && \cite{han2016eie}\\
        CSP&$E_{\text{CSP}}$& $E_\text{LP} + E_\text{CE} + E_\text{PA}$ & & & Num. of bits & $b$ & 12\\
        &&\quad $+ E_{\text{ECSP}} + E_{\text{DAC}}$ & \si{\joule} & & Num. of served users & $K$ \\
         &&&&& Num. of antennas per \gls{csp} & $M$ \\
        \cmidrule[\heavyrulewidth](r){1-5} \cmidrule[\heavyrulewidth](l){6-10}
         \multicolumn{5}{l}{\footnotesize $^\ddag$ based on DELL EMC S4148T-ON (\SI{336}{\watt} / 48 ports)~\cite{9815696}.}& \multicolumn{5}{l}{\footnotesize $^*$ obtained by finding the \gls{pae} at $P_\text{t,max}$ in~\cite{powerAmplifierQorvo}.} \\
        \end{tabularx}
        \vspace{-2mm}
\end{table*}

\subsubsection{Fronthaul}
In this system, an Ethernet-based fronthaul is assumed, where synchronization, data, and control are multiplexed over the same cables. In this network, all \glspl{csp} and \glspl{ecsp} have a \gls{wr} core~\cite{6070148} providing time, phase, and frequency synchronization. \Gls{wr} is a system using well-established IEEE Ethernet-based standards, i.e., timing protocols (\gls{synce} and \gls{ptp}) to distribute time over a network. In addition, it uses phase frequency detectors to measure the fine-grained phase difference between all nodes in the network~\cite{6070148}, thereby attaining picosecond-level synchronization. For this reason, \acrlong{wr} is being actively studied to enable practicable and scalable \gls{cfmmimo} systems~\cite{Bigler2018, callebautTestbeds}.

The energy consumption of the \glspl{ecsp} is entirely defined by this fronthaul, i.e., the Ethernet link and one \gls{wr} core.

\subsubsection{Power Amplifier}
The \glspl{csp} feature a \gls{pa} per \gls{rf} chain. The power consumption of the \gls{pa} is given as
\begin{equation}
P_\text{PA} = \frac{1}{\eta_\text{max}} \left(\frac{P_\text{t,max}}{P_\text{t}}\right)^\beta P_\text{t}, \label{eq:PA-power} 
\end{equation}
where $\eta_\text{max} \in (0,1]$ is the \gls{pae} achieved when the transmit power is $P_\text{t}=P_\text{t,max}$. This parameter and $\beta$ are \gls{pa}-dependent, where a typical value for $\beta$ would be between 0.4 and 0.5~\cite{6515206}.\footnote{In this work, $\beta$ is set to $0.5$ to facilitate optimization with an objective function involving Euclidean norm.}

\subsubsection{Sampling and Processing}
The energy model of the \gls{csp} considered in this work also encompasses the energy consumption associated with sampling and processing (\cref{tab:energy}). 

A 12-bit \gls{adc} sampling at \SI{600}{\mega\hertz} is utilized, which is active during the pilot phase. These samples are down-sampled and processed for channel estimation. A \gls{dac} with the identical parameters are used during the downlink data transmission. The respective energy consumption depends on the energy consumption per arithmetic operation. Here, we assume a two-input multiplier as the typical arithmetic operation, where every operation requires retrieving operands from storage elements and storing them, e.g., register files, on-chip \gls{sram}, or off-chip \gls{dram}.

\subsubsection{Energy Model}
The full energy model for the \glspl{ecsp} and \glspl{csp} for one coherence~block becomes,
\begin{align}
    E_{\text{ECSP}} &= (P_\text{eth}+P_\text{sync})  \frac{\tau_c}{f_\text{BB}}\\
    E_{\text{CSP}}&=E_\text{LP} + E_\text{CE} + E_\text{PA}+ E_{\text{ECSP}} + E_{\text{DAC}}.
\end{align}
Details are summarized in \cref{tab:energy}.

\subsection{System Model}
Let $\mathcal{S}=\{1,\ldots,S\}$ and $\overline{\mathcal{S}}=\{1,\ldots,\overline{S}\}$ denote the sets of \glspl{csp} and \glspl{ecsp}, respectively, where $S=|\mathcal{S}|$ and $\overline{S}=|\overline{\mathcal{S}}|$. Predefined sets of \glspl{csp} are connected to each \gls{ecsp}, forming disjoint sets. We let $\mathcal{S}(\overline{s})$ denote the set of \glspl{csp} connected to \gls{ecsp} $\overline{s}$. We have $\mathcal{S}(\overline{s})\bigcap \mathcal{S}(\overline{s}')=\emptyset$ for $\overline{s}\neq \overline{s}'$ and $\bigcup_{\overline{s}\in \overline{\mathcal{S}}}\mathcal{S}(\overline{s})=\mathcal{S}$.  We assume that each \gls{ue} belongs to one federation and is served jointly by the \glspl{csp} of that federation. For a given snapshot, suppose that there are $K$ \glspl{ue} connected to the network. The set of \glspl{ue} is denoted by $\mathcal{K}=\{1,\ldots,K\}$, where $K=|\mathcal{K}|$. The maximum possible number of federations for joint transmission is $F$ and the set of federations is denoted by $\mathcal{F}=\{1,\ldots,F\}$, where $F=|\mathcal{F}|$. $|\mathcal F|$ should be large enough that there will be enough federations to accommodate all \glspl{ue} and \glspl{csp}, and, in general, not all federations $f \in \mathcal F$ need to be used. A simple upper bound for $|\mathcal F|$ is the number of \glspl{ue}, since one will never need more federations than would be enough to place each \gls{ue} in its own federation, but in practice $|\mathcal F|$ can be much smaller than this.

We assume a block fading channel model, where the channels take independent realizations in each coherence~block. We let $\tau_p$ denote the number of mutually orthogonal pilot sequences. To eliminate pilot contamination, each joint transmission federation can only serve up to $\tau_p$ UEs so that they can share mutually orthogonal pilot sequences. Let $\tau_c>\tau_p$ denote the number of channel uses in each coherence block. The payload data is transmitted using $\tau_c-\tau_p$ channel uses in each coherence~block.

We let $M$ denote the number of antennas per \gls{csp}. We will assume uncorrelated Rayleigh fading and let $h_{k,s,m}\in \mathcal{N}_{\mathbb{C}}(0,\beta_{k,s})$ denote the channel from the $m$th antenna of CSP $s$ to UE $k$, where the large-scale fading channel coefficient is $\beta_{k,s}>0$.
We let $\rho_p$ denote the normalized \gls{snr} of each pilot symbol. The channels $h_{k,s,m}$ can be estimated using the \gls{mmse} estimator~\cite{demir2021foundations} and the variance of the \gls{mmse} estimate of the channels $h_{k,s,m}$ can be computed as 
\begin{align}
    \gamma_{k,s} = \frac{\tau_p\rho_p \beta_{k,s}^2}{\tau_p\rho_p \beta_{k,s}+1}.
\end{align}

We let $x_k^f\in\{0,1\}$ denote the binary variable representing whether UE $k$ belongs to federation $f$ or not, i.e., $x_k^f=1$ if UE $k$ belongs to federation $f$ and $x_k^f=0$ otherwise. Similarly, $y_s^f\in \{0,1\}$ denotes whether CSP $s$ belongs to federation $f$ or not. Note that each CSP and each UE can only belong to one federation, i.e., $\sum_{f\in \mathcal{F}}y_s^f\leq1$, $\forall s$ and $\sum_{f\in \mathcal{F}}x_k^f=1$, $\forall k$. 

Different sets of federations use orthogonal time-frequency resources so that there is no inter-federation interference. We assume that CSP $s$  uses the total transmit power of $(\rho_s^f)^2$ if it belongs to federation $f$ and active, so we have the relation
\begin{align}
    \rho_{s}^f\leq \sqrt{P_{\rm max}}y_s^f. \label{eq:rho-y}
\end{align}

Under this scenario, an achievable downlink data rate of UE $k$ belonging to one of the federations in $\mathcal{F}$ using \gls{mrt} precoding and equal power allocation with $(\rho_s^f)^2/\tau_p$ can be derived from \cite[Corol.~6.3 and 6.4]{demir2021foundations} as\footnote{The \gls{mrt} precoding and equal power allocation enable to express the highly combinatorial problem in a more manageable form.}
\begin{align}
   & R_k^{\rm dl} = \frac{\tau_c-\tau_p}{\tau_c}\nonumber \\
    &\times\log_2\left(1+\frac{M/\tau_p\left(\sum_{f\in\mathcal{F}}\sum_{s\in \mathcal{S}}x_k^{f}\rho_{s}^{f}\sqrt{\gamma_{k,s}}\right)^2}{\sum_{f\in\mathcal{F}}\sum_{s\in\mathcal{S}}x_k^f\left(\rho_{s}^{f}\right)^2\beta_{k,s}+\sigma^2}\right)
\end{align}
where $\sigma^2$ is the noise variance at the receiver of each UE $k$.

 \FloatBarrier

\section{Federation Assignment: Problem Formulation and Proposed Algorithm}\label{sec:fed_assign}
In this section, we formulate the Federation Assignment problem as an optimization problem, discussing its cost and constraints. The solution of such problem is challenging in general, and therefore we propose a tailored \textit{divide-and-conquer} heuristic for it.

\subsection{Problem Formulation}

The overall aim is to minimize total energy consumption under the UE data rate constraints and federation assignment rules. The data rate constraints are $R_k^{\rm dl}\geq R_k^{\rm thr}$,
which are equivalent to
\begin{align}
    \frac{M/\tau_p\left(\sum_{f\in\mathcal{F}}\sum_{s\in \mathcal{S}}x_k^{f}\rho_{s}^{f}\sqrt{\gamma_{k,s}}\right)^2}{\sum_{f\in\mathcal{F}}\sum_{s\in\mathcal{S}}x_k^f\left(\rho_{s}^{f}\right)^2\beta_{k,s}+\sigma^2} \geq \mathrm{SINR}_{k}^{\rm thr}
\end{align}
where $\mathrm{SINR}_{k}^{\rm thr}=2^{R_k^{\rm thr}\tau_c/(\tau_c-\tau_p)}-1$. It can easily be shown that the above constraints are equivalent to $F$ subconstraints given as
\begin{align}
&x_k^fM/\tau_p\left(\sum_{s\in \mathcal{S}}\rho_{s}^{f}\sqrt{\gamma_{k,s}}\right)^2  \geq x_k^f\mathrm{SINR}_{k}^{\rm thr}\nonumber\\
&\quad \times\left( \sum_{s\in\mathcal{S}}\left(\rho_{s}^{f}\right)^2\beta_{k,s}+\sigma^2 \right), \quad   \forall k\in \mathcal{K},\forall f\in \mathcal{F}
\end{align}
which will later ease the management of the optimization problem.
Given a set of UEs and applications running on them, we can formulate the federation resource allocation problem as follows:
\begin{subequations}
    \label{form:fed_assign}
    \begin{align}
       & \underset{\{x_k^f,y_s^f,z_{\overline{s}},\rho_{s}^f\}}{\textrm{minimize}} \quad \overline{\text{E}}_{\text{CSP}} \sum_{s\in \mathcal{S}}\sum_{f\in\mathcal{F}} y_s^f+\text{E}_{\text{ECSP}} \sum_{\overline{s}\in \overline{\mathcal{S}}} z_{\overline{s}}\nonumber\\
       &\quad +(\tau_c-\tau_p)f_\text{BB}^{-1}\frac{\sqrt{P_\text{t,max}}}{\eta_\text{max}}\sum_{s\in \mathcal{S}}\sqrt{\sum_{f\in\mathcal{F}}\left(\rho_{s}^f\right)^2}\label{eq:fa_obj}\\
        &  \textrm{subject to:} \nonumber \\
        & x_k^fM/\tau_p\left(\sum_{s\in \mathcal{S}}\rho_{s}^{f}\sqrt{\gamma_{k,s}}\right)^2  \geq x_k^f\mathrm{SINR}_{k}^{\rm thr}\nonumber\\
&\quad \times\left( \sum_{s\in\mathcal{S}}\left(\rho_{s}^{f}\right)^2\beta_{k,s}+\sigma^2 \right), \quad   \forall k\in \mathcal{K},\forall f\in \mathcal{F} \label{SINR-constraint} \\
        & \rho_{s}^f\leq \sqrt{P_{\rm max}}y_s^f, \quad \forall s\in \mathcal{S}, \forall f\in \mathcal{F} \label{power-constraint}\\
        &  \sum_{f\in \mathcal{F}}y_s^f \leq z_{\overline{s}},  \quad \forall s \in \mathcal{S}(\overline{s}), \forall \overline{s}\in\overline{\mathcal{S}} \label{csp-ecsp-relation} \\
        & \sum_{f \in \mathcal F} x_k^f = 1, \quad  \forall k \in \mathcal K \label{eq:fa_user_only_one}\\
         & \sum_{k \in \mathcal{K}} x_k^f \leq \tau_p,  \quad \forall f \in \mathcal F \label{eq:user-number-limit}\\
        &  x_k^f \in \{0, 1\},  \quad \forall k \in \mathcal K,  \forall f \in \mathcal F \label{eq:binary1}\\
        & y_s^f \in \{0, 1\}, \quad  \forall s \in \mathcal S, \forall f \in \mathcal F \label{eq:binary2} \\
     &    z_{\overline{s}} \in \{0, 1\}, \quad  \forall \overline{s} \in \overline{\mathcal{S}} \label{eq:binary3}
    \end{align}
\end{subequations}
where $\overline{\text{E}}_{\text{CSP}}=\text{E}_{\text{CSP}}-\text{E}_{\text{PA}}$, the objective function in \eqref{eq:fa_obj} is the total energy consumption per coherence block and $\sqrt{\sum_{f\in\mathcal{F}}\left(\rho_{s}^f\right)^2}$ is used in place of $\sqrt{P_\text{t}}$ in \eqref{eq:PA-power}. The SINR constraint for each UE is given in \eqref{SINR-constraint} and the relation between $\rho_s^f$ and $y_s^f$ from \eqref{eq:rho-y} is set in the constraint \eqref{power-constraint}. The constraints in \eqref{csp-ecsp-relation} ensure that ECSP $\overline{s}$ is activated ($z_{\overline{s}}=1$) if at least one of the CSPs connected to it is active. These constraints also guarantee that each active CSP belongs to only one federation, i.e., $\sum_{f\in \mathcal{F}}y_s^f\leq1$. Similarly, the constraints in \eqref{eq:fa_user_only_one} limit the number of federations each UE connects to one. To allow for mutually orthogonal pilot sequences in each federation, the number of maximum UEs in each federation is restricted by $\tau_p$ in \eqref{eq:user-number-limit}. Finally, the constraints in \eqref{eq:binary1}-\eqref{eq:binary3} represent binary constraints.

Solving problem~\eqref{form:fed_assign} is challenging in practice, as it can be seen as a mixed integer program with non-convex constraints. The following section will propose a heuristic that allows to approximate a solution efficiently.

\subsection{Proposed Algorithm}
First, we notice that even if the binary constraints were relaxed, the optimization problem given in~\eqref{form:fed_assign} is hard to solve due to highly coupled optimization variables. Furthermore, its mixed-integer nature adds to the complexity of the problem.
The first step then is to reformulate the problem. In particular, we approximate~\eqref{form:fed_assign} by adding auxiliary variables $\epsilon_s^f\geq 0$, $\tilde{\epsilon}_k^f\geq 0$ and related penalty terms to the objective function as
\begin{subequations}
    \label{form:fed_assignB}
    \begin{align}
       & \underset{\{x_k^f,y_s^f,z_{\overline{s}},\rho_{s}^f,\epsilon_s^f,\tilde{\epsilon}_{k}^f\}}{\textrm{minimize}} \quad \overline{\text{E}}_{\text{CSP}} \sum_{s\in \mathcal{S}}\sum_{f\in\mathcal{F}} y_s^f+\text{E}_{\text{ECSP}} \sum_{\overline{s}\in \overline{\mathcal{S}}} z_{\overline{s}}\nonumber\\
       &\quad +(\tau_c-\tau_p)f_\text{BB}^{-1}\frac{\sqrt{P_\text{t,max}}}{\eta_\text{max}}\sum_{s\in \mathcal{S}}\sqrt{\sum_{f\in\mathcal{F}}\left(\rho_{s}^f\right)^2} \nonumber\\
       &\quad +\lambda\left(\sum_{k\in \mathcal{K}}\sum_{f\in \mathcal{F}}\tilde{\epsilon}_k^f+\sum_{s\in \mathcal{S}}\sum_{f\in \mathcal{F}}\epsilon_s^f\right)\label{eq:fa_objB}\\
        &  \textrm{subject to:} \nonumber \\
        & x_k^f\sqrt{M/\tau_p}\sum_{s\in \mathcal{S}}\rho_{s}^{f}\sqrt{\gamma_{k,s}}+\tilde{\epsilon}_k^f \geq x_k^f\sqrt{\mathrm{SINR}_{k}^{\rm thr}}\nonumber\\
&\quad \times\sqrt{ \sum_{s\in\mathcal{S}}\left(\rho_{s}^{f}\right)^2\beta_{k,s}+\sigma^2 }, \quad   \forall k\in \mathcal{K},\forall f\in \mathcal{F} \label{SINR-constraintB} \\
        & \rho_{s}^f\leq \sqrt{P_{\rm max}}y_s^f+\epsilon_s^f, \quad \forall s\in \mathcal{S}, \forall f\in \mathcal{F} \label{power-constraintB}\\
        & \epsilon_s^f\geq 0, \quad \tilde{\epsilon}_k^f\geq 0, \quad \forall s\in \mathcal{S}, \forall k \in \mathcal{K}, \forall f\in \mathcal{F} \\
        & \eqref{csp-ecsp-relation}-\eqref{eq:binary3}
    \end{align}
\end{subequations}
where $\lambda\gg 1$ is a large penalty coefficient to force $\epsilon_s^f$ and $\tilde{\epsilon}_k^f$ to zero. These slack variables are introduced to relax the constraints and facilitate convergence.

The next step is to apply a \textit{divide-and-conquer} heuristic to~\eqref{form:fed_assignB} based on an \textit{alternating minimization} scheme. The idea is to alternate the solution of two subproblems; in particular, for each iteration $\ell \in \{1, 2, \ldots \}$, we perform the following:
\begin{enumerate}
    \item \textit{Power allocation with fixed assignment}: fixing the binary variables $\{x_k^f,y_s^f,z_{\overline{s}}\}$, we solve~\eqref{form:fed_assignB} for the continuous optimization variables $\{\rho_s^f,\epsilon_s^f,\tilde{\epsilon}_k^f\}$. The resulting problem is a convex programming problem since the constraints in \eqref{SINR-constraintB} can be expressed as second-order cone constraints and the functions in the objective and the other constraints are convex.

    \item \textit{Federation and CSP/ECSP assignment with fixed power}: fixing $\{\rho_s^f\}$ to the solution obtained in 1), \eqref{form:fed_assignB} is now solved for the binary variables $\{x_k^f,y_s^f,z_{\overline{s}}\}$ and slack variables $\{\epsilon_s^f,\tilde{\epsilon}_k^f\}$. The problem in this case is a mixed-integer linear programming problem, and can be solved by branch-and-bound-type algorithms.
\end{enumerate}
The proposed algorithm continues alternating between 1) and 2) until a termination condition is reached.

{\emph {\bf Remark 1:}} The optimization problem under consideration presents greater complexity compared to the AP selection problems previously addressed in the literature \cite{demir2024, chien2020}. This heightened complexity arises from the inclusion of additional binary variables that correspond to the federation assignment of APs and UEs. To tackle this challenge, we propose a new solution that involves a novel mathematical manipulation, which enables decoupling of the binary and continuous optimization variables.

After extensive simulations, we observed that randomly activating CSPs until the SINR-constrained power allocation becomes feasible performs better when the number of CSPs is relatively small. However, when a relatively high number of active CSPs are required, the random activation method struggles to find a solution, while our proposed algorithm succeeds. To leverage the strengths of both approaches, we add a random activation refinement stage at the end of the proposed algorithm.

\section{Evaluation in an Industrial Factory Environment}\label{sec:evaluation}
The proposed algorithm is evaluated in an industrial environment. The carrier frequency is \SI{3}{GHz}. The path loss, i.e., large-scale fading, is simulated based on the channel models proposed by ETSI and 3GPP~\cite{ETSImodels}, more specially the \textit{indoor factory (sparse clutter, high base station)} is used.
The probability that a \gls{ue} is in \gls{los} follows the model proposed in~\cite{ETSImodels}. A fixed number of \glspl{ecsp} is selected, i.e., \num{5} \glspl{ecsp}.

The simulated industrial factory hall measures 12 meters in width, 20 meters in length, and 10 meters in height. The \gls{csp} units are evenly distributed and integrated into the ceiling. Additionally, $K=24$ \glspl{ue} are simulated at random locations within the hall.

\definecolor{CSP15}{HTML}{1b9e77}
\definecolor{CSP30}{HTML}{d95f02}
\definecolor{CSP60}{HTML}{7570b3}

\pgfplotsset{default_ls/.style={solid, mark=*}}
\pgfplotsset{M8/.style={dotted, mark=pentagon*}}
\pgfplotsset{M16/.style={dashed, mark=square*}}
\pgfplotsset{M32/.style={dashdotted, mark=diamond*}}

\begin{figure}
    \centering
\begin{tikzpicture}[spy using outlines={rectangle, magnification=2, connect spies}]

    \begin{axis}[
        ylabel={Consumed Power (\si{\watt})},
        xlabel={Data Rate (\si{\mega\bit\per\second})},
        width=0.7\linewidth, 
        minor x tick num=3,
        minor y tick num=4,
        legend columns=3,
        cycle list/Dark2,
        legend style={nodes={scale=0.8, transform shape}},
        legend style={at={(0.5,1.02)}, anchor=south},
        grid=major,
         xmin=10, xmax=96,
        ymin=-9.9, ymax=520.0, 
    ]
    \pgfplotstableread{results/powers-v4.txt}\myenergydata;

    \addlegendimage{empty legend}\addlegendentry{15 CSPs}
    \addlegendimage{empty legend}\addlegendentry{30 CSPs}
    \addlegendimage{empty legend}\addlegendentry{60 CSPs}
     
    \addplot[CSP15,default_ls] table [x=data-rate, y=factory-15-M32] {\myenergydata};
    \addlegendentry{M32}

    \addplot[CSP30,default_ls] table [x=data-rate, y=factory-30-M16] {\myenergydata};
    \addlegendentry{M16}
    
    \addplot[CSP60,default_ls] table [x=data-rate, y=factory-60-M8] {\myenergydata};
    \addlegendentry{M8}

     \addlegendimage{empty legend}\addlegendentry{}

     \addplot[CSP30, M32] table [x=data-rate, y=factory-30-M32] {\myenergydata};
    \addlegendentry{M32}
     
     \addplot[CSP60, M32] table [x=data-rate, y=factory-60-M16] {\myenergydata};
    \addlegendentry{M16}

    \addplot[CSP15, M8] table [x=data-rate, y=factory-15-M8] {\myenergydata};
    \addlegendentry{M8}






      


    \end{axis}
    
    \end{tikzpicture}
    \vspace{-2mm}
    \caption{Total power consumption in terms of data rate for different numbers of CSPs and antennas. Each row in the legend corresponds with the same number of total antennas.}
    \label{fig:results}
      \vspace{-2mm}
\end{figure}
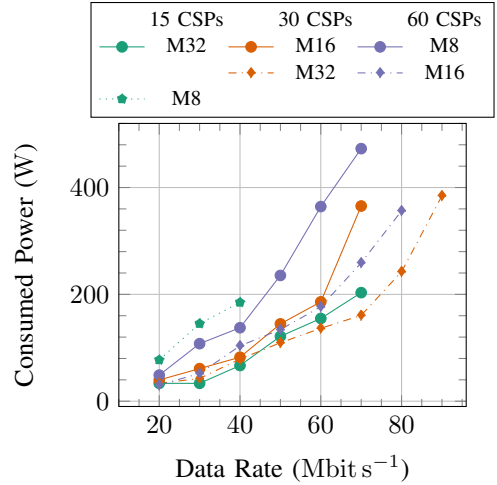







    

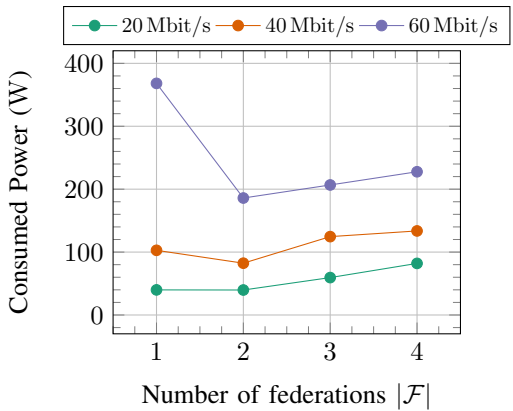
\begin{figure}
    \centering
\begin{tikzpicture}[spy using outlines={rectangle, magnification=2, connect spies}]

    \begin{axis}[
        ylabel={Consumed Power (\si{\watt})},
        xlabel={Number of federations $|\mathcal{F}|$},
        width=0.7\linewidth, 
        minor x tick num=3,
        minor y tick num=4,
        legend columns=3,
        cycle list/Dark2,
        legend style={nodes={scale=0.8, transform shape}},
        legend style={at={(0.5,1.02)},anchor=south},
        grid=major,
         xmin=0.5, xmax=4.5,
        ymin=-29.9, ymax=420.0,
    ]
    \pgfplotstableread{results/power-federations-v3.txt}\myactivedata;

    \addplot[CSP15, default_ls] table [x=federations, y=factory-20Mbit] {\myactivedata};
    \addlegendentry{\SI{20}{Mbit/s}}

    \addplot[CSP30, default_ls] table [x=federations, y=factory-40Mbit] {\myactivedata};
    \addlegendentry{\SI{40}{Mbit/s}}

    \addplot[CSP60, default_ls] table [x=federations, y=factory-60Mbit] {\myactivedata};
    \addlegendentry{\SI{60}{Mbit/s}}

    \end{axis}
    
    \end{tikzpicture}
      \vspace{-2mm}
    \caption{Power consumption in terms of number of federations for different rate requirements.}
      \vspace{-2mm}
    \label{fig:results2}
\end{figure}

In Fig.~\ref{fig:results}, we set the number of federations to $F=2$ and the pilot channel uses to $\tau_p=K/F=12$, resulting in UEs being forced to be in two federations. We consider several numbers of CSPs and antennas per CSPs, $M$. If the problem is infeasible, no point is shown in the figure for the corresponding data rate requirement. For example, 15 CSPs with $M=8$ antennas  cannot provide 50\,Mbit/s and more. When we increase the number of antennas per CSP for a given number of CSPs, we can both improve the feasibility and reduce power consumption. Moreover, this figure showcases that less distributed \gls{cfmmimo} structure provides better energy saving. This effect was not observed in original works on cell-free massive MIMO, as they only focus on radio power consumption. However, since the cost per CSP is much larger than the number of antennas, we observe this effect.

In Fig.~\ref{fig:results2}, we set $\tau_p=K/F$ and vary $F$. There are $S=30$ CSPs with $M=16$. We observe that $F=2$ federations are optimal for power consumption when the data rate requirement is low. As the data rate requirement increases, the advantages of using $F=2$ federations for energy savings become more pronounced. Fig.~\ref{fig:results2} illustrates the versatile nature of the proposed system, showcasing a significant reduction in power consumption when the requested downlink data transfer rate is low. However, the system has the capability to scale up its performance as needed, depending on the real-time load.

\glsresetall 
\section{Conclusion}\label{sec:conclusion}

This study highlights that in certain scenarios, a co-located deployment exhibits energy advantages over a \gls{cf} network. This preference stems from the increased energy overhead associated with a substantial number of \glspl{csp}. The implication is that exclusively considering transmit power consumption might lead to different conclusions than when examining the complete energy outlay of the system.

An intriguing avenue for future research involves delving into the energy distribution among distinct system elements and assessing their behaviors as the system scales. Additionally, we advocate for a paradigm shift, emphasizing the importance of not solely optimizing the energy or spectral efficiency of systems. Instead, the focus should be on addressing real-time user requirements and curbing energy consumption to meet these needs precisely—no more, potentially yielding outcomes distinct from optimization solely based on factors like spectral efficiency.

\section*{Acknowledgement}
The authors express their gratitude to Ove Edfors and Fredrik Tufvesson for their insightful feedback and support, and Liang Liu and Baktash Behmanesh for their valuable expertise regarding the utilized energy model.

\printbibliography

\end{document}